\documentclass[a4paper,10pt,twoside,twocolumn]{article}
\usepackage{ercoftac_2012}

\begin{document}
\setcounter{page}{11}

\title{{\vspace{-0.0cm}\sc{Manufactured Turbulence with Langevin equations}}}
\author{Aashwin Ananda Mishra$^1$, Sharath Girimaji$^2$\vspace{0.3cm}\\
\small{\it{$^1$Aerospace Engineering Department, Texas A\&M University ,College Station, Texas, USA}}\\
\small{\it{$^2$Aerospace Engineering Department, Texas A\&M University ,College Station, Texas, USA}}\\}
\date{}
%\mail{kuban@imc.pcz.czest.pl}
\maketitle
\thispagestyle{fancyplain}
\vspace{0.5cm}

\section*{Abstract}
By definition, \emph{Manufactured turbulence}(MT) is purported to mimic physical turbulence rather than model it. The MT equations are constrained to be simple to solve and provide an inexpensive surrogate to Navier-Stokes based Direct Numerical Simulations (DNS) for use in engineering applications or theoretical analyses. In this article, we investigate one approach in which the linear inviscid aspects of MT are derived from a linear approximation of the Navier-Stokes equations while the non-linear and viscous physics are approximated via stochastic modeling. The ensuing Langevin MT equations are used to compute planar, quadratic turbulent flows. While much work needs to be done, the preliminary results appear promising.

\section{Introduction}

Turbulence is an enigmatic mix of \emph{method} (large scale coherent structures) and \emph{madness} (chaotic, small scale motions). While the coherent structures are evidently flow-dependent, the small-scale chaotic motions exhibit a rather surprising level of independence from the large scales (Kolmogorov hypotheses). Arguably, it is the large scale structures that are dynamically important and the role of the small scale motions is merely to provide a means for dissipating the cascaded energy. It is rather interesting that the dynamically decisive large scales are easier to compute and more difficult to model than the small scales which are more onerous to compute but play a more straightforward role.  Any attempt at capturing turbulence physics must pay heed to these crucial matters. 

Our charge in this work is to develop simple-to-solve equations that mimic physical turbulence, rather than model it. Here we reserve the term \emph{model} to indicate those attempts to develop closure equations for the moments of the turbulence field. To \emph{mimic} is to yield spatio-temporal realizations of velocity and pressure fields and entire probability distribution functions. We call such a surrogate flow field, \emph{Manufactured Turbulence}(MT). The MT flow-field is intended for use in engineering applications and theoretical analyses as an inexpensive substitute to the Direct Numerical Simulations (DNS) of the Navier-Stokes equations.

In the absence of an analytical theory of turbulence, the computational recourse to turbulence is extensively utilized in industrial and academic applications. Of these, computationally intensive methods like Direct Numerical Simulation and Large Eddy Simulation are limited in their application due to their excessive computational demands. On the other hand, modeling intensive approaches, such as one or two equation models, are encumbered due to their lack of fidelity in many varieties of flows. In this vein, synthetic or manufactured turbulence is a contrivance to generate signals that mimic real turbulent flow fields. Kinematic Simulation (KS) is predominantly used to this end. 

An alternative that is popular in the turbulent combustion community is based on the Langevin equation in a Lagrangian framework. Such \emph{Probability Density Function methods} have been extensively applied and have become established in turbulence research (\cite{pope1, pope2}). This work is based, in essence, upon extensions of the simplistic analogy between the motion of fluid elements in a turbulent flow and the motion of gas molecules. Chung (\cite{chung}) used a similar analogy with the motion of fluid elements and Brownian motion, to develop a simplified statistical model for turbulence. Kuznetsov and Frost (\cite{kuz}) applied a consonant similitude to use a Langevin equation for this purpose. This was extended by Pope and co workers (\cite{pope3}). In an analogy with the Langevin equation governing the velocity of a particle undergoing Brownian motion, a linear Markov model for fluid particle velocity was developed in \cite{pope3}. The effects of fluctuating pressure and viscosity are modeled via deterministic drift and diffusion terms. The diffusion term represents a random walk in velocity space. Haworth and Pope (\cite{pope3}) used the Navier Stokes equation as the starting point for the model formulation, thus adding physical significance to the terms of the Langevin equation and the concomitant coefficients therein. Furthermore, to account for the rapid component of pressure (and specifically, its dependence on mean gradients) an anisotropic drift term was added to the generic Langevin equation. 

In this article, we apply a general set of Langevin equations to generate Manufactured Turbulence. It is accepted that linear physics provides a qualitative representation for many features of turbulent flows. However, the exactitude of this linear representation is contingent upon many factors. It is found that the quantitative preponderance of linear theory is highly dependent on the regime of flow. This is explained with respect to the nature of the instabilities manifested in these flows.   

\section{Mathematical formulation and rationale}

The essential components of a turbulent flow field consist of:
\begin{enumerate}
\item Linear effects, consisting of inertial physics, embodied in production and \emph{rapid} pressure action.
\item Non-linear effects, that include the \emph{slow} pressure action.
\item Viscous effects. 
\end{enumerate}
Of these, the linear effects are the drivers of turbulence and engender the variations in different flows. Thus, it is essential to ensure that these are represented as precisely as possible. The non-linear effects are universal and can be modeled statistically. Based on physics, pressure action can be decomposed into two components, viz. rapid and slow. 
\begin{equation}
\frac{1}{\rho}\nabla^{2}p'=-2\frac{\partial U_i}{x_j}\frac{\partial u_j}{x_i} - \frac{\partial^2}{\partial x_i\partial x_j}(u_iu_j-\overline{u_iu_j}),
\end{equation}  
where the first and second terms on the right, represent the contributions of rapid and slow pressure, respectively. The adjectives rapid and slow refer to the components of pressure arising, respectively, from the linear and non-linear parts of the source term in the Poisson equation for pressure. The slow component acts to conserve the incompressibility of the velocity field generated by the nonlinear interactions among velocity fluctuations. Similarly, it is the function of rapid pressure to impose the divergence free condition on the fluctuating velocity field produced by linear interactions
between the mean and fluctuating fields. 

Based on established theory, surrogates for the linear and the non-linear effects of pressure can be developed separately. Thence, these can be appended to give a complete, general surrogate for the pressure effects. In contrast to the slow pressure and its universal nature, the action of the rapid
pressure effects are a strong function of the mean velocity field and initial flow conditions. In spite
of the apparent simplification afforded by linearity, the action of rapid pressure is not straightforward. Depending on the nature of the mean velocity field and initial conditions of the flow field, the effect of the rapid pressure component can be diametric. Furthermore, this action can alter the fundamental nature of the flow. This is best exhibited in the regime of elliptic flows, where it is established that the rapid pressure effects initiate and sustain the elliptic flow instability (\cite{camby2}). Most engineering models do not capture the nature of this action and predict a decay of turbulence, contrary to theory and DNS results (\cite{bns}). Thus, the linear pressure effects must be represented as accurately as possible. The import of fidelity to linear dynamics, even in KS has been accepted and attempts have been made to coalesce the knowledge developed via RDT in KS. Nicolleau and Vassilicos (\cite{nw1}) utilized temporal evolution predicted by RDT with the KS velocity field formalism. This was applied and compared contra DNS in \cite{camby3}. Kaneda and Ishida (\cite{ki}) used a similar approach to study the diffusion of a passive scalar. Subsequently, this approach of amalgamating RDT with KS has been extended, for instance in \cite{nw2}. Under the aegis of RDT, the velocity field can be expressed as a summation of advected Fourier modes. In this formulation, the rapid pressure effects can be represented exactly. To this end, the rapid pressure component of the Langevin set is formulated in spectral space. In spectral space, this formulation can account for the initial conditions accurately and is not hampered by an incomplete basis. The germane equations in this regard are:
\begin{equation}\label{kapevo}
\frac{d{{\kappa }_{l}}}{dt}=-{{\kappa }_{j}}\frac{\partial {{U}_{j}}}{\partial {{x}_{l}}},
\end{equation}

\begin{equation}											 
\frac{d{{u}_{j}}}{dt}=-{{u}_{k}}\frac{\partial {{U}_{l}}}{\partial {{x}_{k}}}( {{\delta }_{jl}}-2\frac{{{\kappa }_{j}}{{\kappa}_{l}}}{{{\kappa }^{2}}}),							
\end{equation}
and the incompressibility constraint is given by $\mathbf{u}\cdot\boldsymbol{\kappa}=0$. Herein, $\vec{u}$ and $\vec{\kappa}$, or the Fourier velocity amplitude and wave-vector respectively, are considered random variables and are simulated via Monte Carlo techniques.

With regard to the slow component of pressure, it is established that this has a return to isotropy effect, wherein, the anisotropy of the Reynolds stress tensor is reduced. This, in essence, is a redistribution of the turbulent kinetic energy from any given distribution to an uniform, isotropic distribution. Thus, the slow pressure effects are represented via a stochastic diffusion form. Explicitly,
\begin{equation}
A_{ij}(u,e)dW_j,
\end{equation} where $A_{ij}$ is the diffusion tensor and $dW$ is an isotropic Wiener process. Consequently, the representation reduces to 
\begin{equation}
de_i=g_i(u,e) +A_{ij}(u,e)dW_j+B_{ij}(u,e)dW'_j.
\end{equation}
\begin{equation}
du_i=h_i(u,e) +H_{ij}(u,e)dW_j+G_{ij}(u,e)dW'_j.
\end{equation}
Constraints are applied to the system to ensure physical fidelity. These are:
\begin{enumerate}
\item Ensure that $\vec{e}$ remains a unit vector.
\item Maintain orthogonality of $\vec{u}$ and $\vec{e}$.
\item The PDF of the velocity approaches an isotropic, joint-normal distribution.
\item The evolution of the turbulent kinetic energy is exact in the limit of decaying turbulence.
\end{enumerate}
These ensure realizability of the Reynolds stresses. For details of the derivation, the reader is referred to \cite{vanslooten}.
The velocity evolution equation is represented as a Langevin equation with an anisotropic drift term. Such surrogates can be thought of as bridging methods between one-point closures and multi-point/spectral closures. 
 
The dissipation model is appended to the formulation to complete the basis. This is of the established form:
\begin{equation}
\frac{d{{\epsilon}}}{dt}=\frac{\epsilon^2}{k}(C_1\frac{P}{\epsilon}-C_2).
\end{equation}

Consequently, the entire set of equations reduces to:
\begin{equation}
\begin{split}
du_i=&-u_k\frac{\partial {{U}_{l}}}{\partial {{x}_{k}}}(\delta_{il}-2e_ie_l)dt-\frac{1}{2}\frac{\epsilon}{k}(1+\frac{3}{2}a_u)dt \\ &+\frac{\gamma \epsilon}{k}(b_{ij}-II_b\delta_{ij})u_jdt 
     -\sqrt{a_u \epsilon}dW_i.
\end{split}
\end{equation}
\begin{equation}
\begin{split}     
de_i=&-\frac{\partial {{U}_{m}}}{\partial {{x}_{l}}}e_m(\delta_{il}-e_ie_l)dt-\frac{1}{2}\frac{\epsilon}{k}(a_e+a_u\frac{k}{u_s u_s})e_idt \\ &-\frac{\gamma \epsilon}{k}(\delta_{ij}-2e_ie_j)b_{jl}e_l-\sqrt{a_u \epsilon}\frac{u_i e_l}{u_s u_s}dW_l + \\ & \sqrt{\frac{a_e \epsilon}{k}}(\delta_{il}-e_ie_l-\frac{u_i u_l}{u_s u_s})dW'_l.
\end{split}
\end{equation}
Figure 1 exhibits the representation's performance, wherein the predictions are compared against DNS results (\cite{vanslooten}).

\begin{figure}[h]
\centering
 \includegraphics[width=0.35\textwidth]{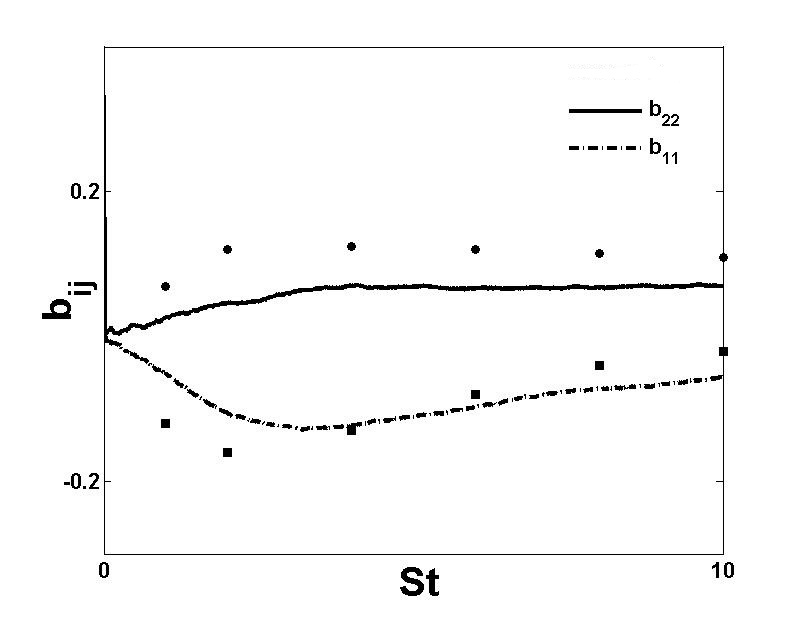}
 \caption{Comparison of the predictions against DNS results.}
 \label{fig:01}
\end{figure}

\section{Linear physics in planar, quadratic flows}

Linear theories such as RDT ignore the interaction of turbulence with itself. This is justified via assumptions regarding the times scales (of mean and fluctuating distortions), a weak turbulence assumption, etc. However, the linear instabilities manifested in RDT obviate these assumptions. With increase in the turbulent kinetic energy, the non-linear effects become more important and thus, linear theory cannot suffice, beyond a very limited time period. In this duration, the linear effects structure the flow field. Thence, non-linear effects modify the evolution of turbulence. This structuring effect of the linear physics is most evident in the instabilities manifested therein, where certain modes are engendered to grow preferentially. Figure 2 exhibits the unstable modes, in a representative hyperbolic and an elliptic flow, with respect to their alignment. The figure is motivated by a congruous illustration in \cite{camby2}.

\begin{figure}[h!]
\centering 
 \includegraphics[width=0.35\textwidth]{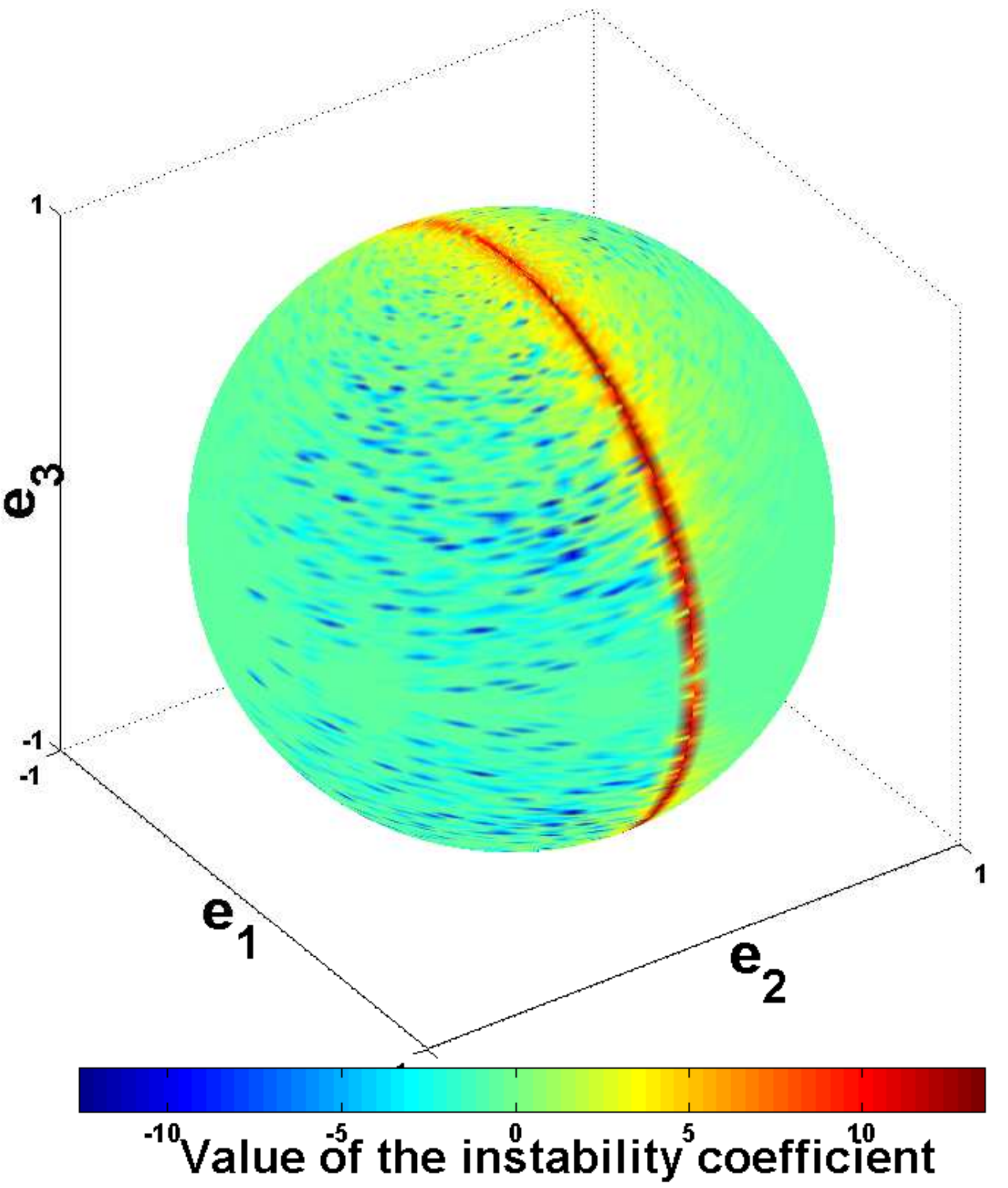}
 \includegraphics[width=0.35\textwidth]{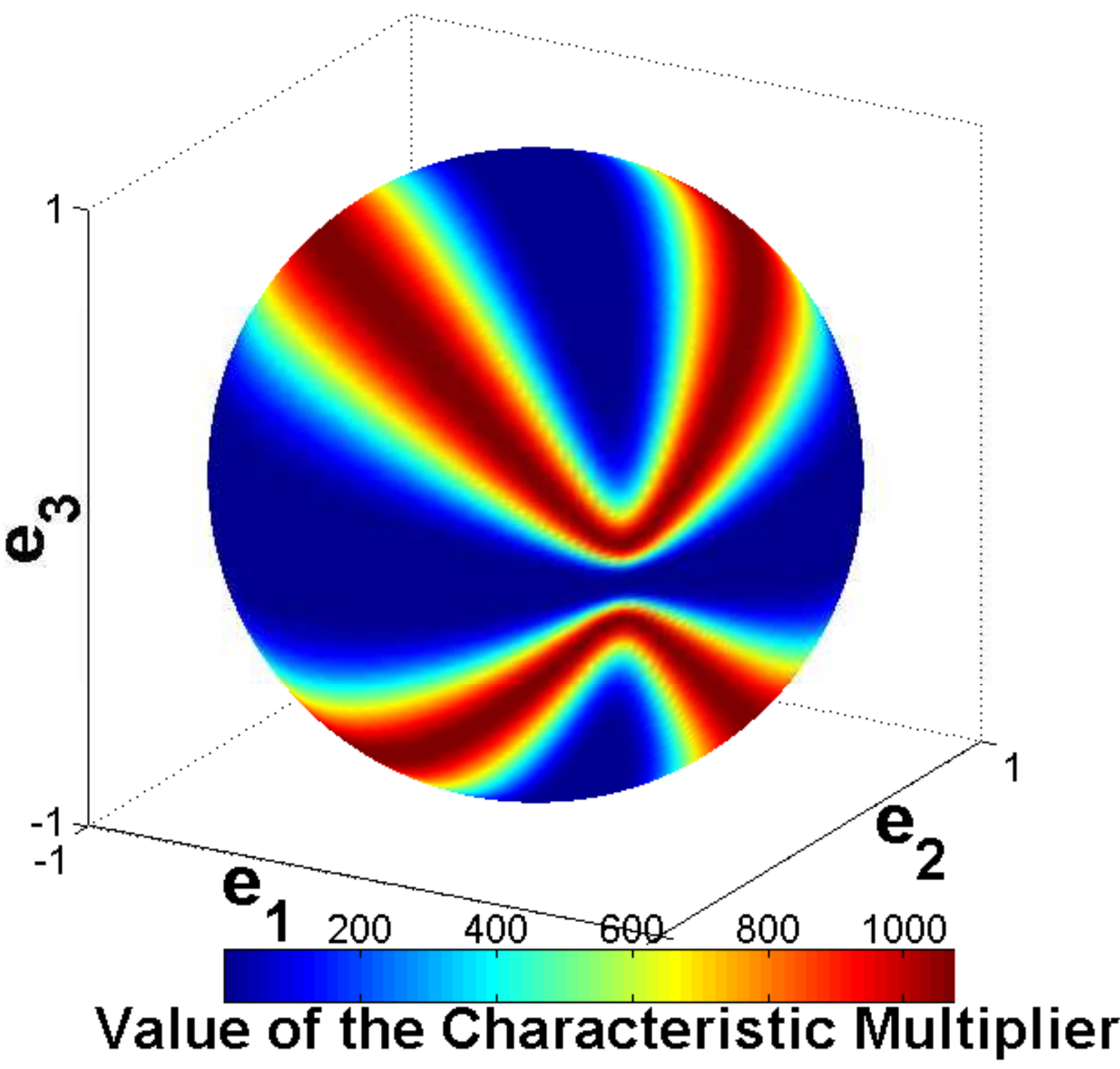}
 \caption{The unstable modes exhibited, with respect to their alignment in (a) a representative hyperbolic flow, (b) in an elliptic flow.}
 \label{fig:01}
\end{figure}

 As can be observed, the unstable modes in an elliptic flow form a continuous band. However, the unstable modes in a hyperbolic flow lie on a set of zero measure. In the hyperbolic case, all other modes are either stable or can undergo some transient growth. Furthermore, this state of alignment for the unstable modes is in itself unstable and these can be forced off this alignment by any perturbations. This is evident in figure 3, wherein the hyperbolic flow instability is arrested by the pressure effects. This occurs via the transfer of turbulent kinetic energy out of the plane of applied shear via the pressure strain correlation. The interested reader is referred to \cite{mg}, wherein the linear aspects of this problem are analyzed in detail.

\begin{figure}[h!]
\centering
 \includegraphics[width=0.35\textwidth]{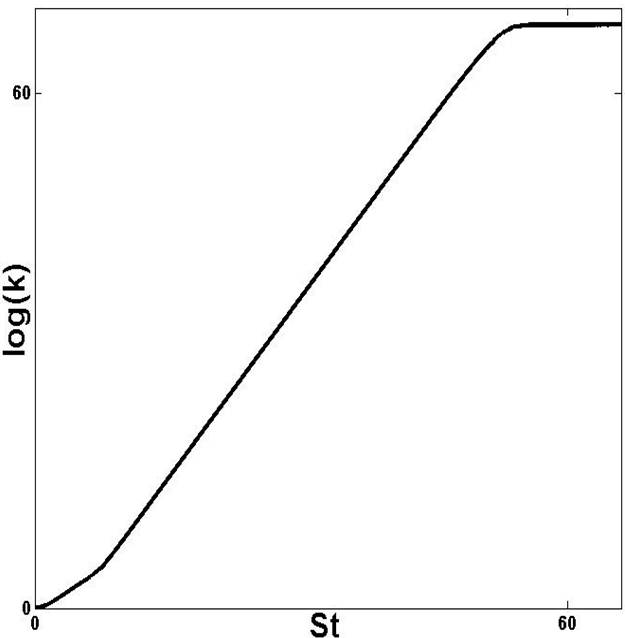}
 \includegraphics[width=0.35\textwidth]{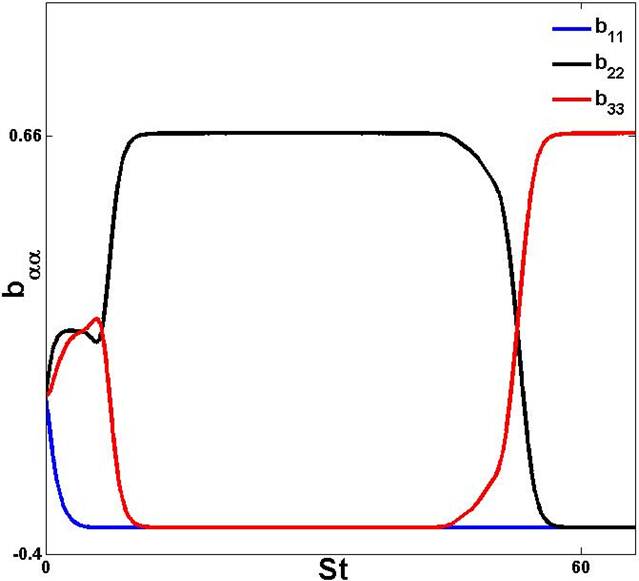}
 \caption{The evolution of (a) the turbulent kinetic energy, (b) Reynolds stress anisotropies in a plane strain flow, under the aegis of the Rapid Distortion Limit.}
 \label{fig:01}
\end{figure}

It is observed that this shift is robust and manifests itself for all open streamline flows, as exhibited in figure 4. In this vein, it is pertinent to question the exactitude of the hyperbolic instability, caused by these modes, in regimes where the non-linear effects become more and more significant. Furthermore, this is contrasted against similar comparisons in other regimes of planar, quadratic flows.

\begin{figure}[h!]
\centering
 \includegraphics[width=0.4\textwidth]{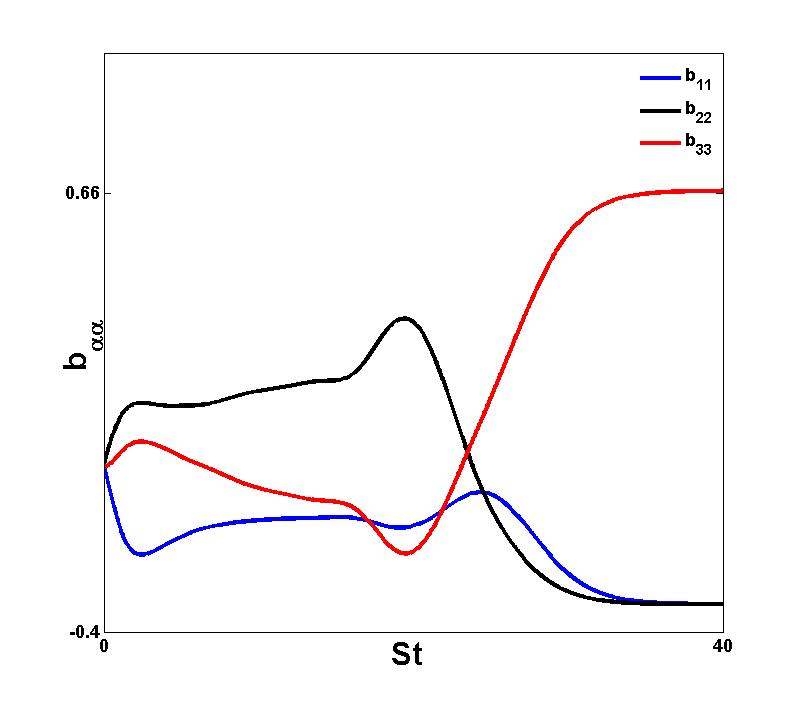}
 \includegraphics[width=0.4\textwidth]{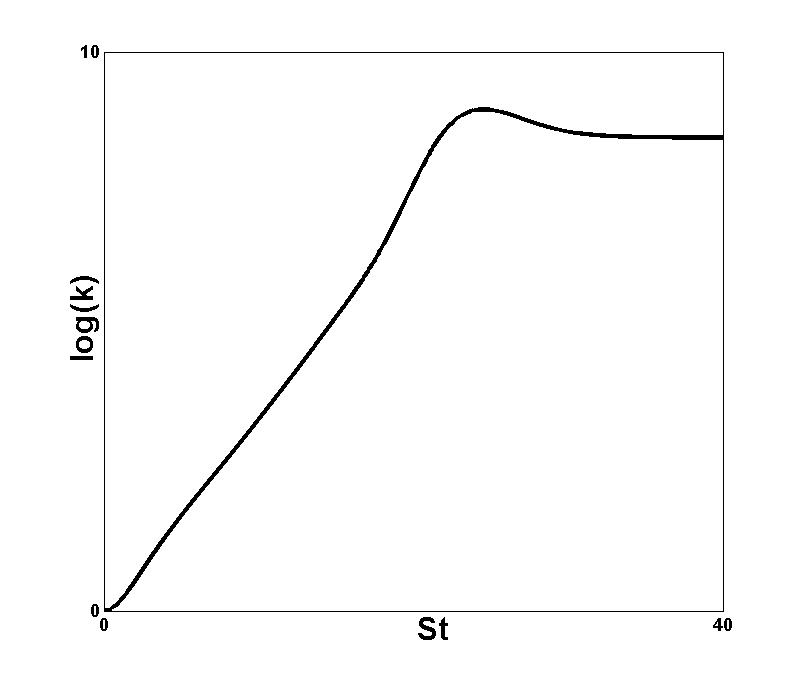}
 \caption{The evolution of (a) the turbulent kinetic energy, (b) Reynolds stress anisotropies in a representative open streamline flow, under the aegis of the Rapid Distortion Limit.}
 \label{fig:02}
\end{figure}

\begin{figure}[h!]
\centering
 \includegraphics[width=0.35\textwidth]{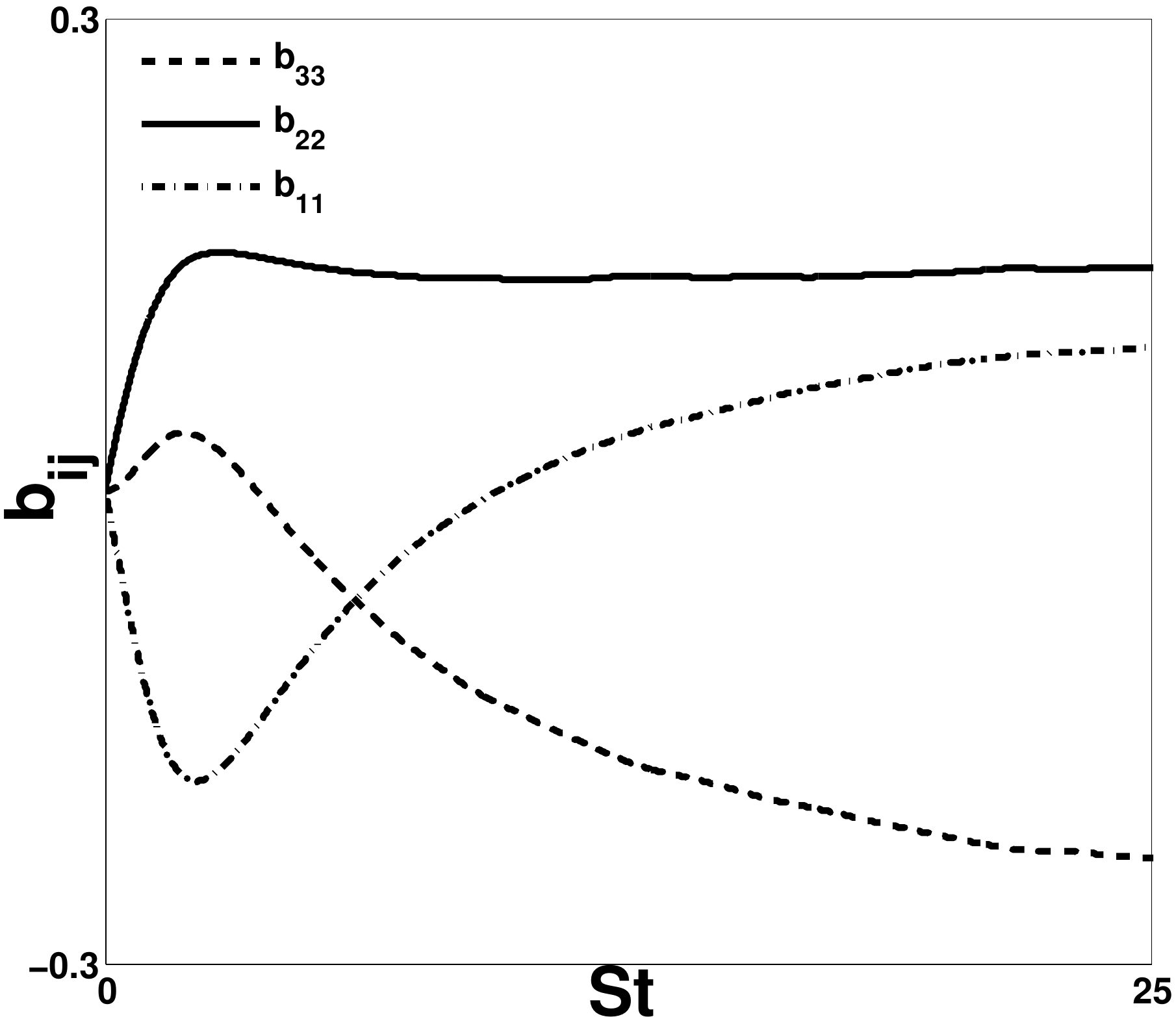}
 \includegraphics[width=0.35\textwidth]{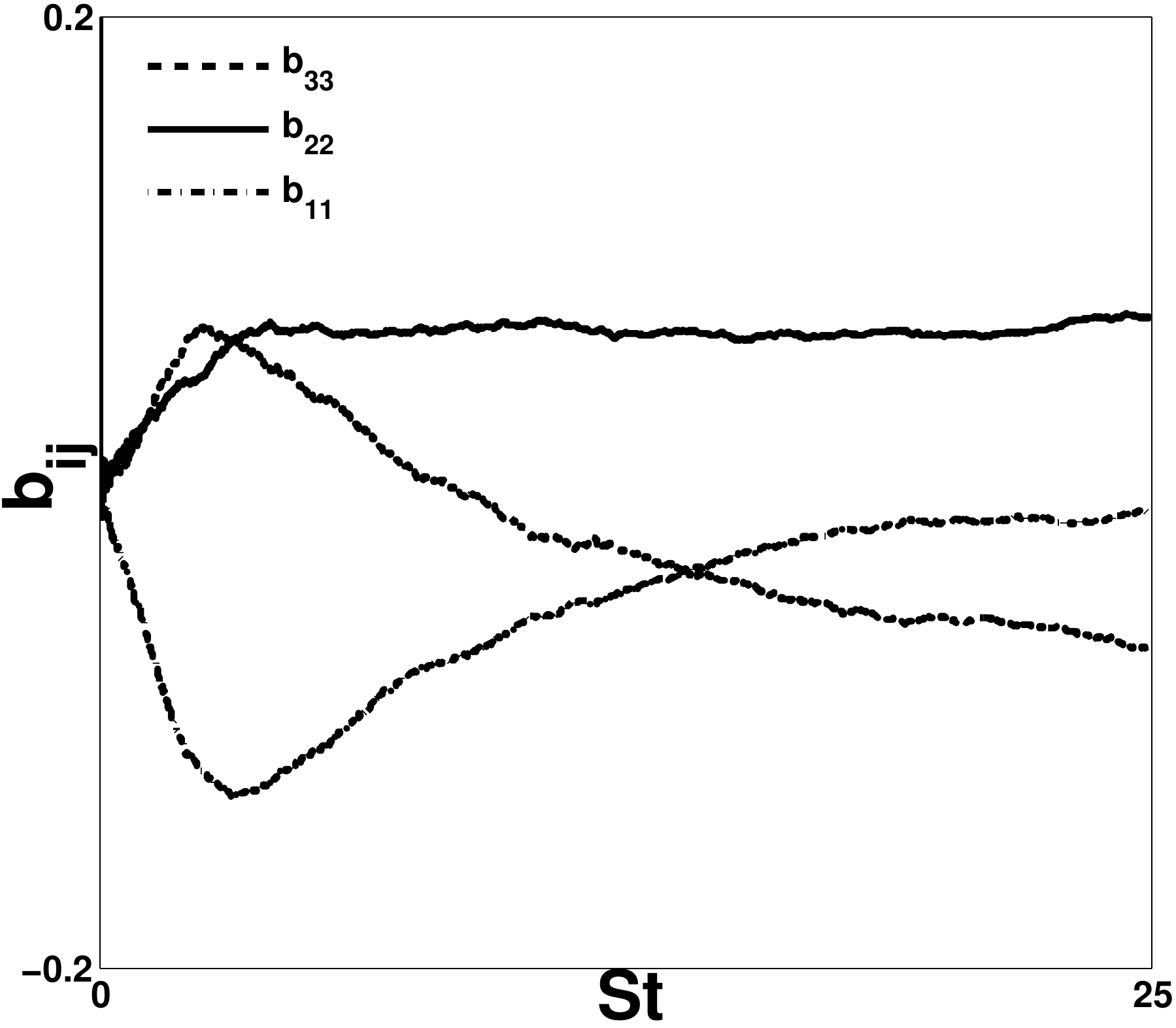}
 \caption{Comparison of the evolution of Reynolds stress anisotropies in a purely sheared flow (a) RDT results, (b) Langevin representation with $\frac{Sk}{\epsilon}=25$. }
 \label{fig:01}
\end{figure}

The structuring effects of linear physics are most predominant in purely sheared flows. This is evident the large streamwise length scales observed in boundary layers. Furthermore, it has been observed that the evolution of flow statistics is similar in DNS studies, as compared to RDT simulations (\cite{camby}). This is exhibited in figure 5, where the results of the Langevin equation representation are compared to those from RDT based simulations. 

\begin{figure}[h!]
\centering
 \includegraphics[width=0.35\textwidth]{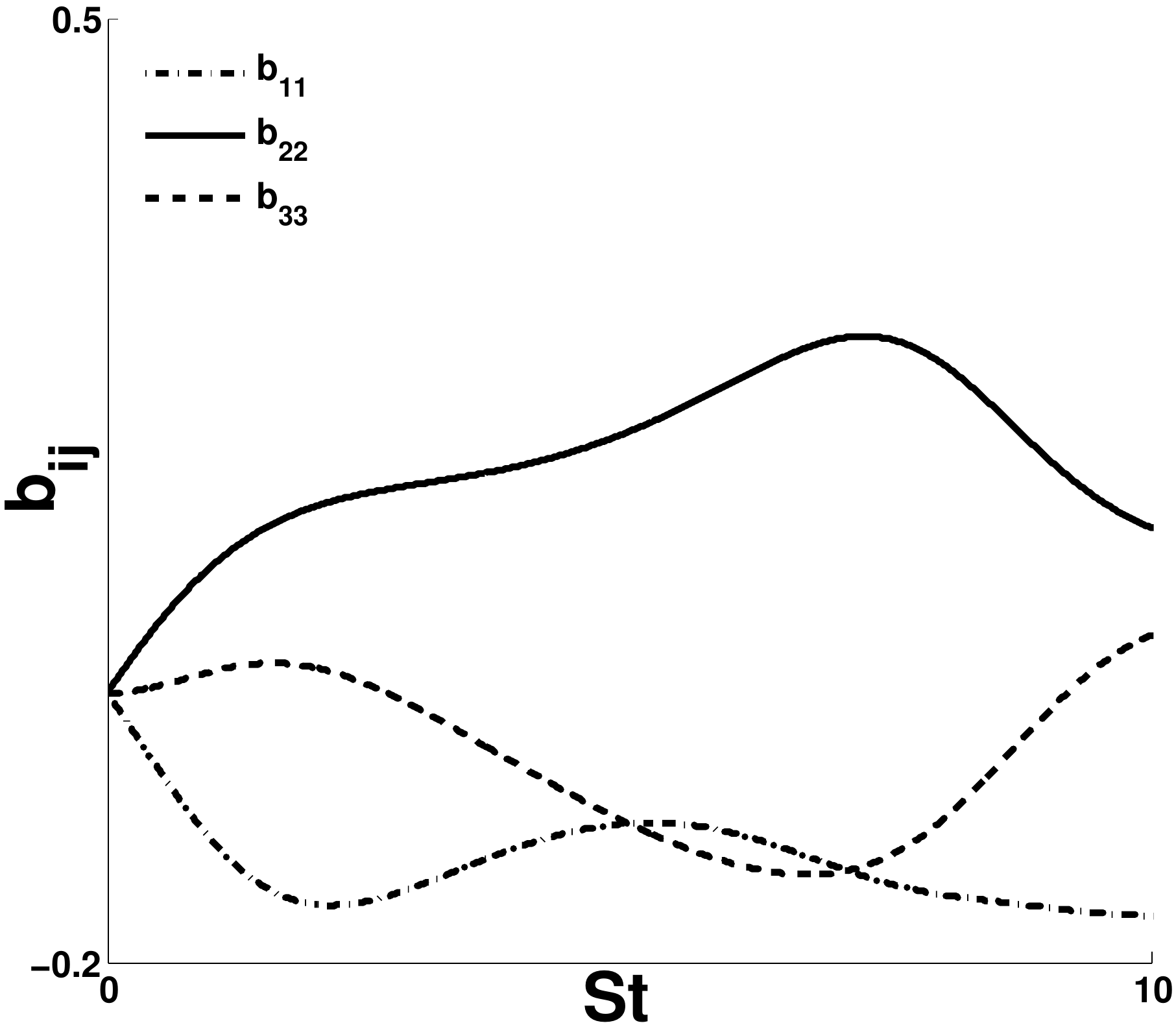}
 \includegraphics[width=0.35\textwidth]{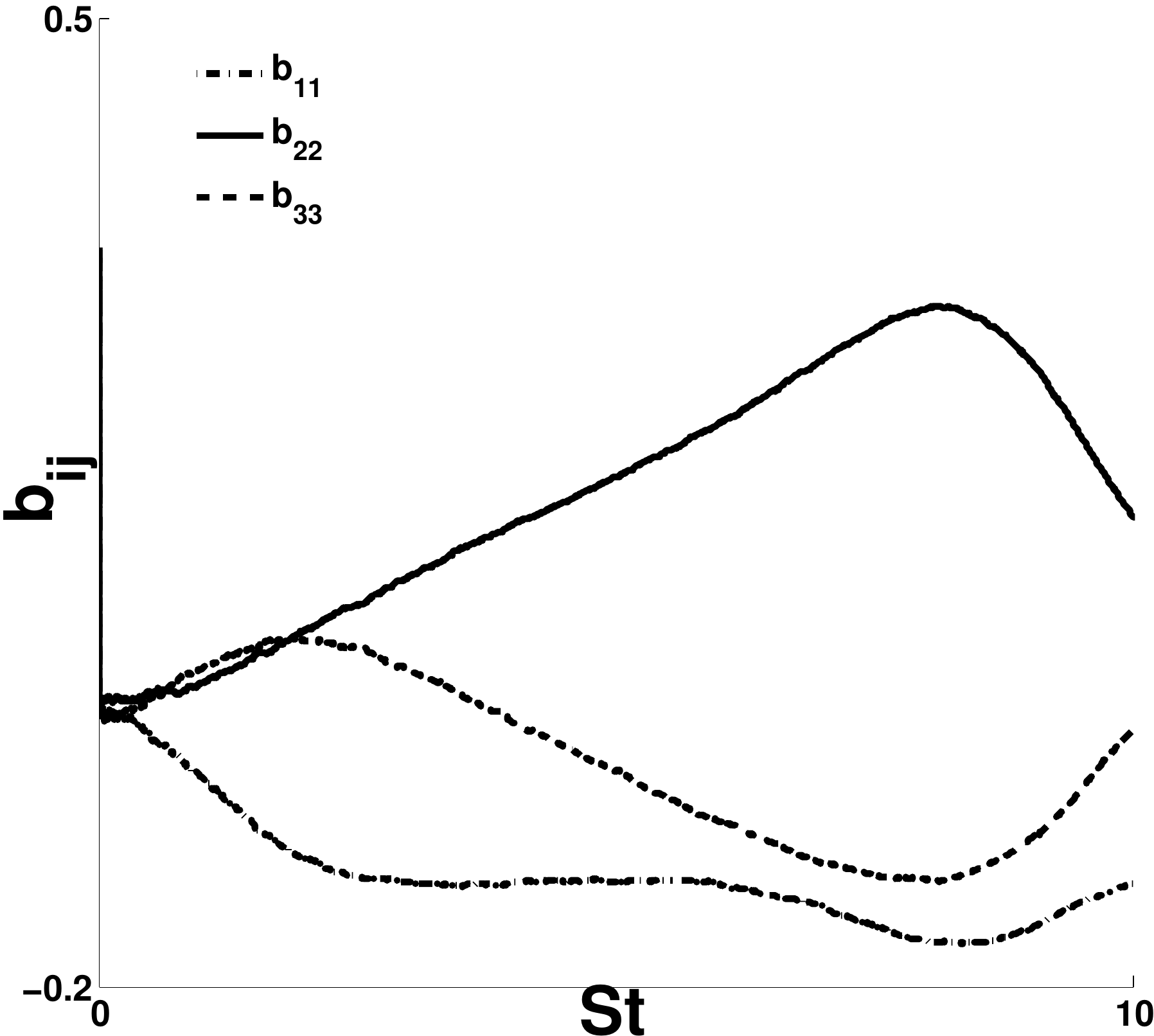}
 \caption{Comparison of the evolution of Reynolds stress anisotropies in a representative elliptic flow (a) RDT results, (b) Langevin representation with $\frac{Sk}{\epsilon}=50$.}
 \label{fig:01}
\end{figure}

Figure 6 compares the evolution of flow statistics for elliptic flows in the presence and absence of non-linear effects. As can be observed, the results are very similar in the absence of non-linear effects or when they are of a small finite value. This is due to the finite measure of the set of unstable modes. However, this scenario does not persist for all elliptic flows. For instance, in purely rotating flows, it is known that linear theory is inconsistent with DNS results (\cite{camby}). 

\begin{figure}[h!]
\centering
 \includegraphics[width=0.45\textwidth]{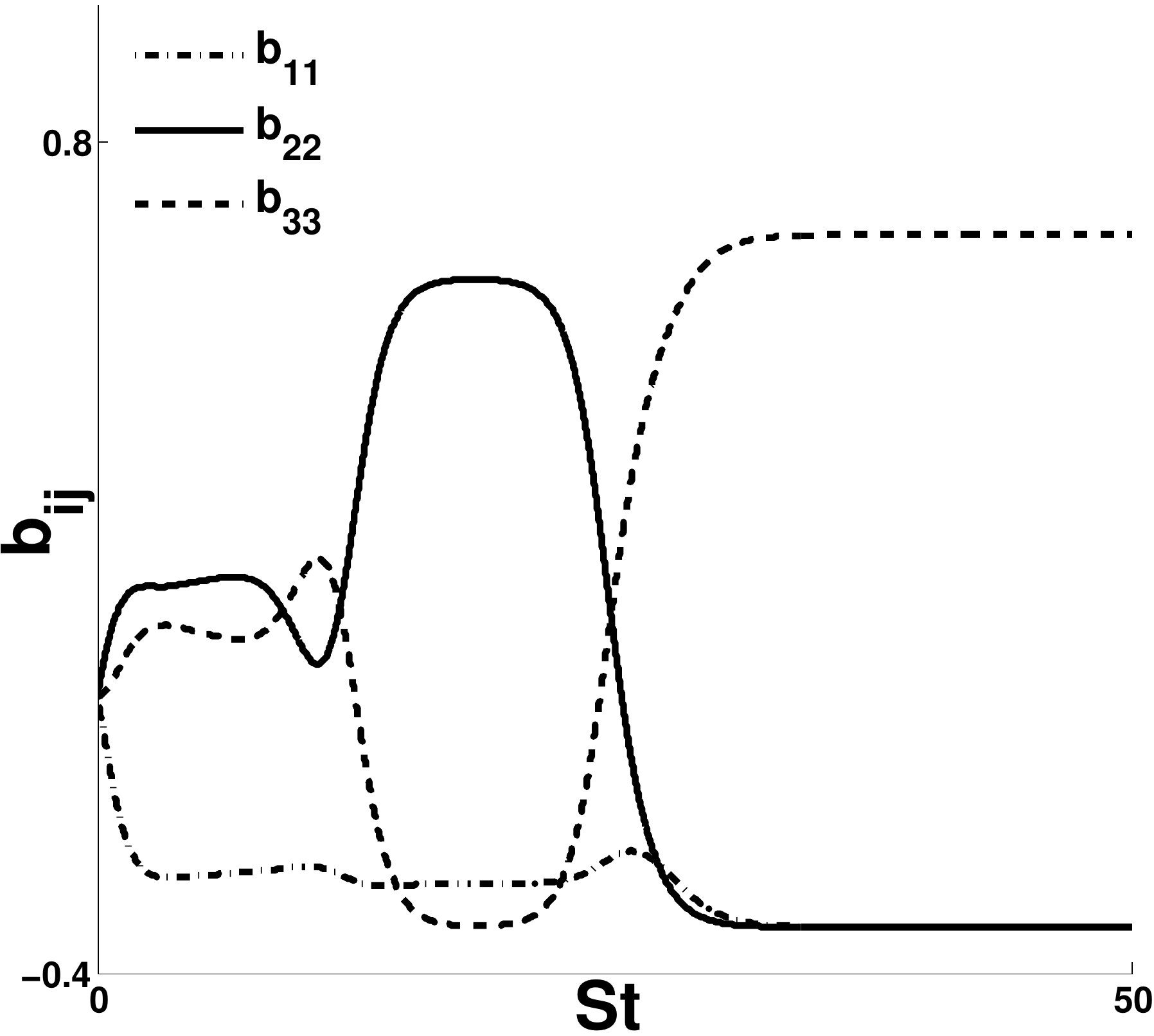}
 \includegraphics[width=0.45\textwidth]{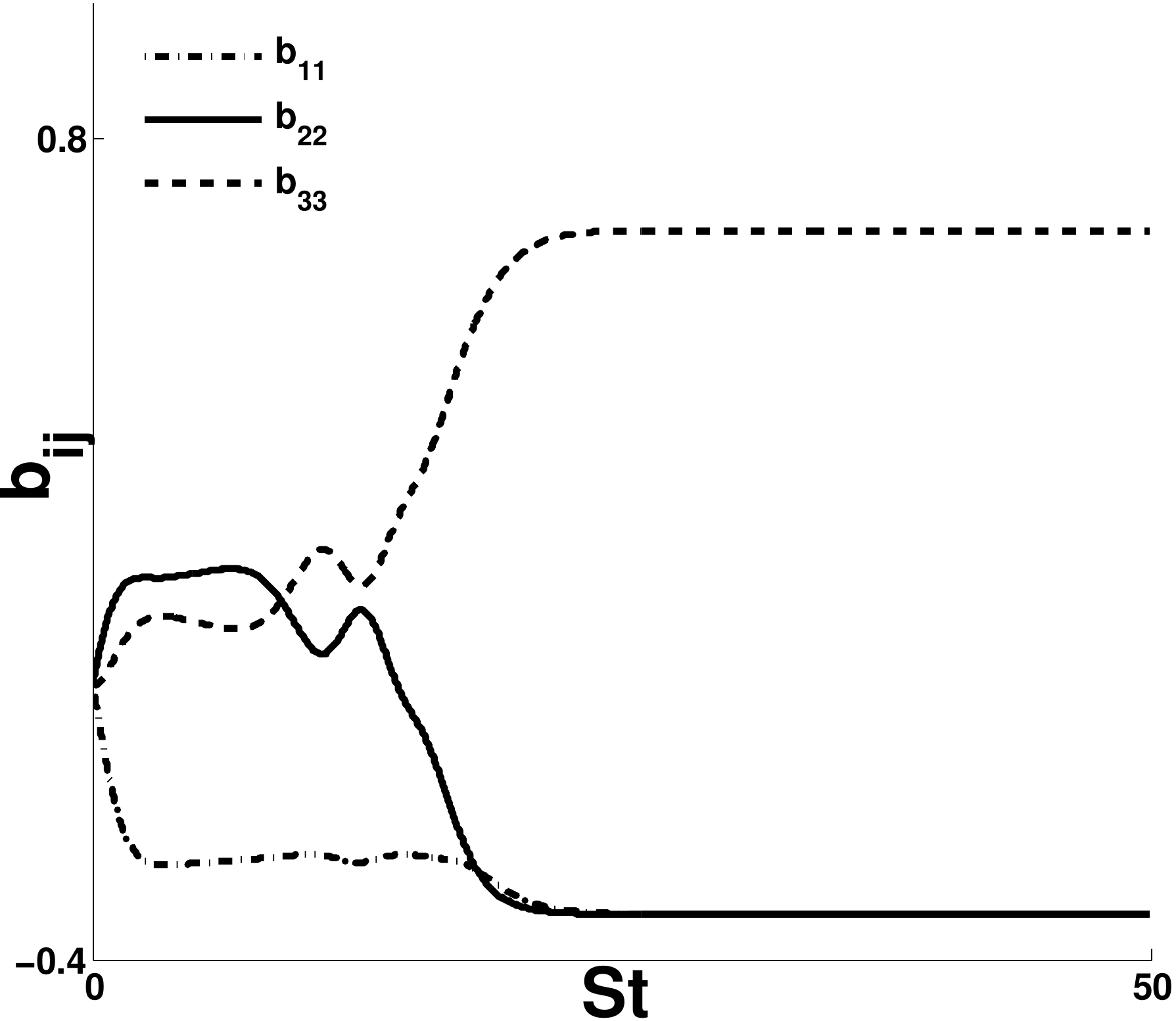}
 \includegraphics[width=0.45\textwidth]{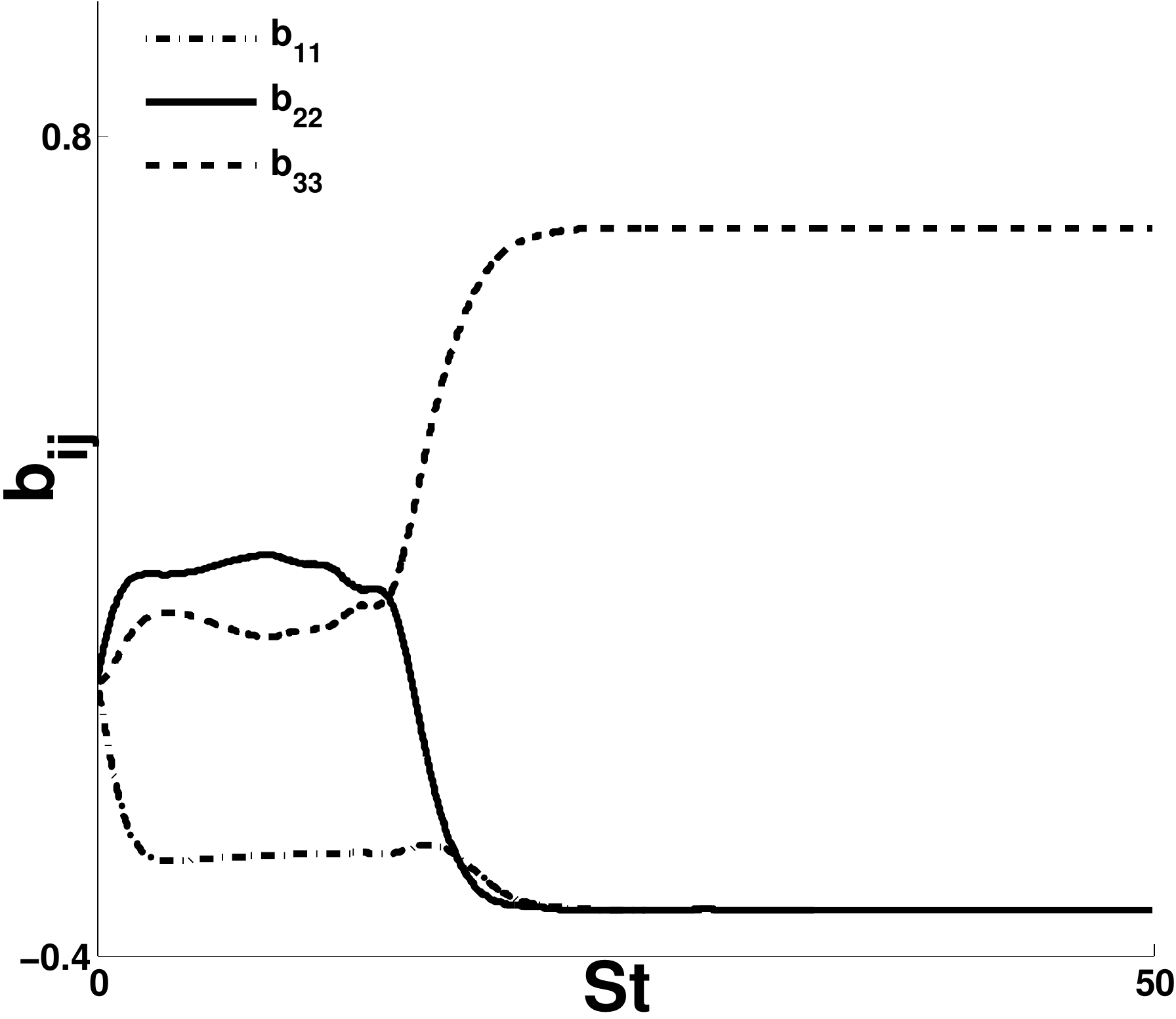}
 \caption{Comparison of the evolution of Reynolds stress anisotropies in a representative hyperbolic flow (a) RDT results, (b) Langevin representation with $\frac{Sk}{\epsilon}=80$, (c) Langevin representation with $\frac{Sk}{\epsilon}=60$.}
 \label{fig:01}
\end{figure}

Figure 7 compares the evolution of flow statistics in a representative hyperbolic flow as the non-linear effects become more important. It is observed that due to the non-linear effects, the switch in the anisotropy evolution occurs progressively earlier. This is due to the perturbation of the wave-vector due to the non-linear effects, which force modes off the unstable set.

\section{Conclusions}
In this article, we exhibit the application of \emph{Manufactured Turbulence} (MT) to study the linear physics in a planar quadratic flow. The MT equations are exact in the Rapid Distortion Limit and use a Langevin equation to simulate the return to isotropy effect of the slow pressure term. Thus, chaotic advection is incorporated using a white noise term. The mathematical formulation of such representations is introduced and the underlying rationale explained.

Thence, this surrogate is applied to study the import of linear physics for planar, quadratic flows. It is found that for purely sheared flows, linear theory provides a very good representation of the evolution of flow statistics, even in the presence of non-linear effects. For general elliptic flows, effects of linear physics are predominant even in the presence of moderate non-linearity. This is due to the banded nature of the instability, where unstable modes lie on a continuous band of finite measure. Thus, perturbations due to the non-linear effects have very little influence on the instability. However, for hyperbolic flows, the linearly unstable modes lie on a set of very small measure. Thus perturbations to these modal alignments may have significant effects on the state of instability and consequently, the evolution of flow statistics. However, only the transient time to reach the asymptotic stage is affected. But the final asymptotic behavior is still as dictated by linear phenomenon. It is observed that linear effects dominate the overall flow behavior, although non-linear aspects can have an important effect on transients.

%Bibliography section! 
\bibliography{references} % corresponds to references.bib file name
\bibliographystyle{ieeetr}
\end{document}